\documentclass[]{aa} 
\usepackage{graphicx}
\begin{document}
   \title{A Substellar Mass Function for Alpha Per\thanks{Based
on observations collected with the Kitt Peak National Observatory, USA,
and the Hispano-German Observatory of Calar Alto, Spain}}


   \author{David Barrado y Navascu\'es
          \inst{1}
          \and
          Jer\^ome Bouvier
          \inst{2}
          \and
          John R. Stauffer
          \inst{3}
          \and  Nicolas Lodieu       \inst{4}
          \and  Mark J. McCaughrean     \inst{4}
          }

   \offprints{D. Barrado y Navascu\'es}

   \institute{Laboratorio de Astrof\'{\i}sica Espacial y F\'{\i}sica 
        Fundamental,
           INTA,  P.O. Box 50727, E-28080 Madrid, Spain\\
              \email{barrado@laeff.esa.es}
         \and Laboratoire d'Astrophysique, Observatoire de Grenoble, 
            Universit\'{e} Joseph Fourier, B.P.  53, 38041
            Grenoble Cedex 9,            France
         \email{jerome.bouvier@obs.ujf-grenoble.fr}
         \and
         IPAC,  California Institute of Technology,
         Pasadena, CA 91125, USA            
         \email{stauffer@ipac.caltech.edu}
        \and
	Astrophysikalisches Institut Potsdam, 
         An der Sternwarte 16, D-14482 Potsdam, Germany
        \email{nlodieu@aip.de,mjm@aip.de}
             }

   \date{Received  ; accepted        }

   \abstract{We present a deep, wide-field optical
 survey of the young stellar cluster
Alpha Per, in which we have discovered a large population of candidate
brown dwarfs. Subsequent infrared photometric follow-up shows that the
majority of them are probable or possible members of the cluster, 
reaching to a minimum mass of 0.035 M$_\odot$. 
We have used this list of 
members to derive the luminosity and mass functions of
the substellar population of the cluster ($\alpha$=0.59$\pm$0.05, 
when expressed
in the mass spectrum form $\phi$$\propto$$M^{-\alpha}$)
and compared 
its slope to the value measure for  the Pleiades. This comparison indicates
that the two cluster mass functions are, indeed, very similar.
   }
   \maketitle
%


\section{Introduction}

In an ongoing effort to discover low-mass stars and brown dwarfs (BDs)
belonging to  young open clusters, we have studied the association
Alpha Per. This is well-known nearby cluster with
(m-M)$_0$=6.23 (176 pc). The interstellar reddening is also low,  A$_V$=0.30
 (Pinsonneault et al. 1998). The normally quoted age for the cluster, based on isochrone
fitting of the upper main sequence (MS), is of order 50 Myr
(cf. Meynet et al. 1993), though models with a larger amount
of convective core overshoot can yield ages up to about
80 Myr (Ventura et al. 1998).
Recently, using the data we published in a preliminary study
of the BD population of the cluster
(Stauffer et al. 1999), we estimated the age as 90 Myr, based on the
location of the lithium depletion boundary.
The theoretical background to this method can be
found in  Kumar (1963) and  
D'Antona and Mazzitelli (1994), and when 
applied to the Pleiades (Basri et al. 1996, Rebolo et al. 1996; 
Stauffer et al. 1998a) and IC\,2391 (Barrado y Navascu\'es et al. 1999)
 has also yielded ages $\sim$50\% older than previously assumed. 
A review of these results may be found in Basri (2000). 

During the last years, different clusters and star forming regions 
have been studied 
intensively  and their BD population revealed.
The Pleiades, Alpha Per cluster, IC2391, M35, NGC2516, Taurus, 
the Trapezium cluster, Sigma Orionis cluster,
 Cha I dark cloud, Upper-Sco OB association, 
and IC348  have been investigated 
(Rebolo et al. 1995; 
Festin 1997; 
Bouvier et al. 1998;
Stauffer et al. 1998ab, 1999;
Brice\~no et al. 1998;
Barrado y Navascu\'es et al. 1999, 2001ab, 2002;
Neuh\"auser \& Comeron 1999;
Zapatero-Osorio et al. 1999, 2000; 
Lucas \& Roche 2000;
Luhman 1999, 2000;
Mart\'{\i}n et al. 1999, 2000, 2001;
B\'ejar et al. 2001; 
Pinfield et al. 2000;
Ardila et al. 2000; 
Najita et al. 2000;
Moraux et al. 2001;
{\it et cetera}).
 All these works show that BDs are  
quite numerous and that the mass function (MF) 
usually presents an increase for very low-mass objects.
In any case, the total mass below the substellar limit 
only contributes a few percent to the total mass
 of the parent cluster, at least in the case of the 
Pleiades  ($\sim$3-5 \%, Bouvier et al. 1998, Hodgkin \& Jameson 2000).
However, it is not clear whether this MF is universal and this result
can be extrapolated to other young clusters.

In this paper, we present a new deep, wide-field
optical survey of the Alpha Per cluster. We have followed-up the optical
candidates in the near-infrared using new infrared imaging data and the
2MASS catalogue (Skrutskie et al. 1997). Using this wealth of data, we
have been able to establish the presence of a substantial population of
BDs in the cluster, and derived its substellar MF.


\section{Observations}

\subsection{Optical survey}

%
%
   \begin{figure}
   \centering
   \includegraphics[width=8cm]{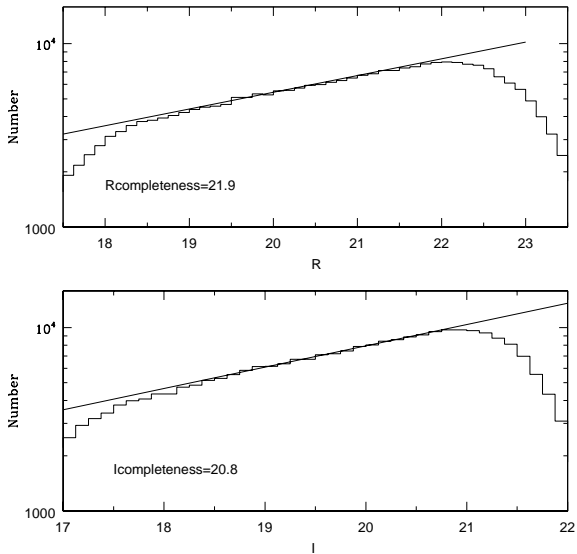}
      \caption{Completeness of our KPNO/MOSA survey.
 For candidate members of the 
Alpha Per cluster, the survey is complete down to $I_c$=19.5,
 ($R$-$I$)$_c$=2.4}
         \label{Fig1}
   \end{figure}
%

\setcounter{table}{0}
\begin{table}
\caption[]{Field centers for KPNO 4m CCD Mosaic Images}
\begin{tabular}{lcc}
\hline
 Field   &    R.A.        &  DEC         \\
\cline{2-3}
         &  \multicolumn{2}{c}{(2000.0)} \\
\hline				       
    A    &     03:27:10.0 & 49:24:00.0   \\
    B    &     03:28:00.0 & 48:42:00.0   \\
    C    &     03:24:50.0 & 48:42:00.0   \\
    D    &     03:21:40.0 & 48:42:00.0   \\
    E    &     03:19:35.0 & 49:27:00.0   \\
    F    &     03:33:45.0 & 49:35:00.0   \\
    G    &     03:33:45.0 & 49:04:00.0   \\
    H    &     03:33:45.0 & 48:34:00.0   \\
    I    &     03:33:45.0 & 50:06:00.0   \\
    J    &     03:30:30.0 & 50:09:00.0   \\
    K    &     03:27:00.0 & 50:13:00.0   \\
    L    &     03:23:10.0 & 48:11:00.0   \\
    M    &     03:21:00.0 & 50:48:00.0   \\
\hline
\end{tabular}
$\,$
\end{table}

We obtained deep optical imaging of the Alpha Per
open cluster using the KPNO 4-m CCD mosaic camera, MOSA, on November
19--22 1998. Field centres were selected in
order to cover as much of the central part of the
cluster as possible while avoiding the brightest stars
in the field (Table 1).  All of the images were obtained
with the facility $R$ and $I$ filters.  Weather during the run
was variable:  Nov. 19 was generally overcast with periods
of only moderate cirrus; Nov. 20 was clear but with 
extremely poor seeing; Nov. 21 was clear with good seeing;
and Nov. 22 was again partly cloudy.  In order to provide
calibrated photometry for all stars, we overlapped the
survey fields so as to provide photometric calibration
even for fields whose deep images were obtained through
light cirrus.  The average seeing for the deep images
was of order 1.0 arcseconds.   For most fields, we 
obtained two 900 second
$R$ images and two 420 second $I$ images; a few of the fields
just designed to tie together the photometry were only
observed with 100 or 200 second exposures.  In all,
we obtained deep imaging of 13 fields, thus corresponding
to a total area on the sky of order 3 square degrees.  
Photometric calibration was derived from short exposures of 
three Landolt (1992) fields in SA92, SA98 and SA101.  
Initial results from analysis of the shorter exposures
obtained during this run were reported by Stauffer
et al. (1999).

%
%
   \begin{figure}
   \centering
   \includegraphics[width=8cm]{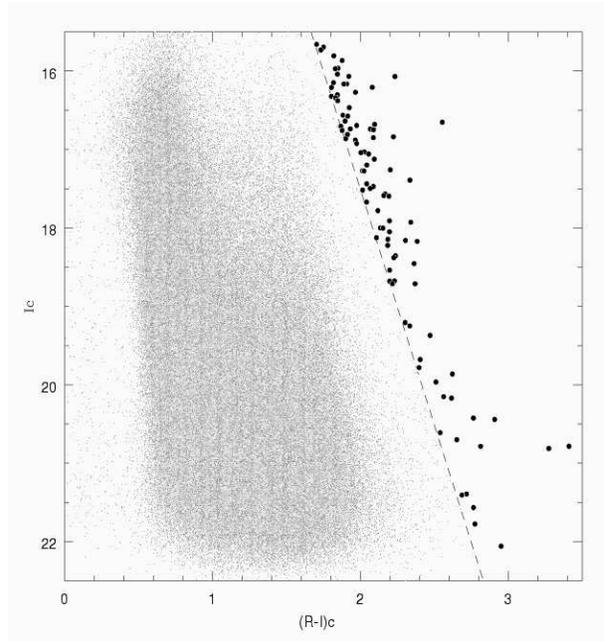}
      \caption{Colour--Magnitude diagram in the fields we have studied.
        Solid circles represent the location of our initial selection
of cluster candidates. The dashed line corresponds to the fiducial
cluster main sequence.}
         \label{Fig2}
   \end{figure}
%

\subsection{Infrared follow-up}

In addition to the KPNO/MOSA observations, we 
 undertook several infrared observing campaigns
at the Hispano-German Calar Alto Observatory in order to obtain additional 
data for candidate cluster members (see section 3.2). These runs included November
22--23 1999, using Omega-Cass (1024$^2$ pixels, 0.3 arcsec/pixel) on the
3.5-m telescope; February 16--21 2001 and November 6--7 2001, using 
MAGIC (256$^2$ pixels, 0.64 arcsec/pixel) on the 2.2-m telescope; and
December 11--14 2000 and December 28--31 2001 using Omega-Prime (1024$^2$ 
pixels, 0.4 arcsec/pixel) on the 3.5-m telescope. We observed a total of 
53 Alpha Per candidates in the magnitude range $K$$^\prime$=13.0--19.2 magnitudes,
equivalent to $I$$_c$=15.7--21.7 magnitudes. All were observed in the $K$$^\prime$ filter,
while 24 were also measured at $J$, and a handful also at $H$.

%
%
   \begin{figure*}
   \centering
   \includegraphics[width=8cm]{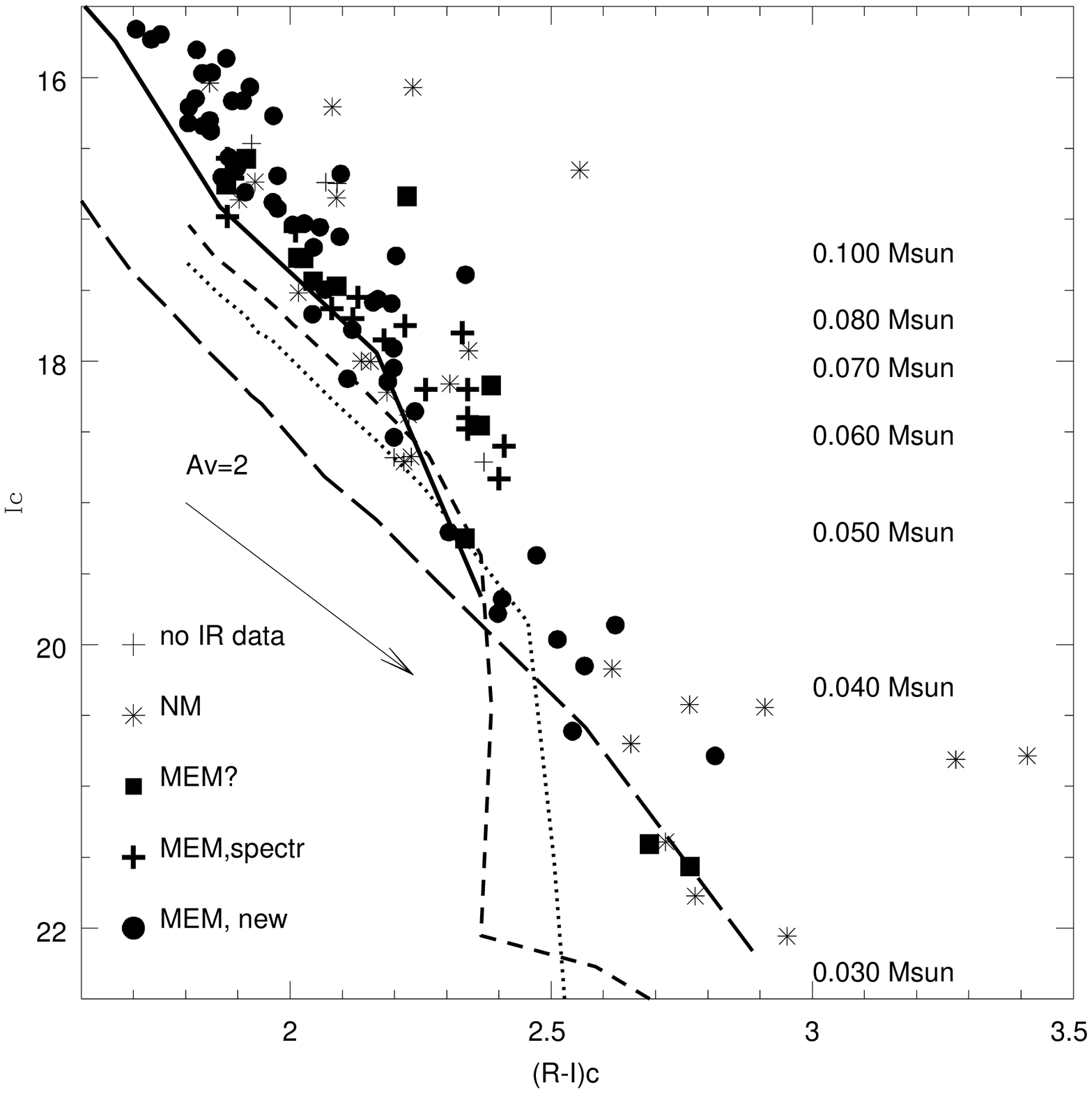}
   \includegraphics[width=8cm]{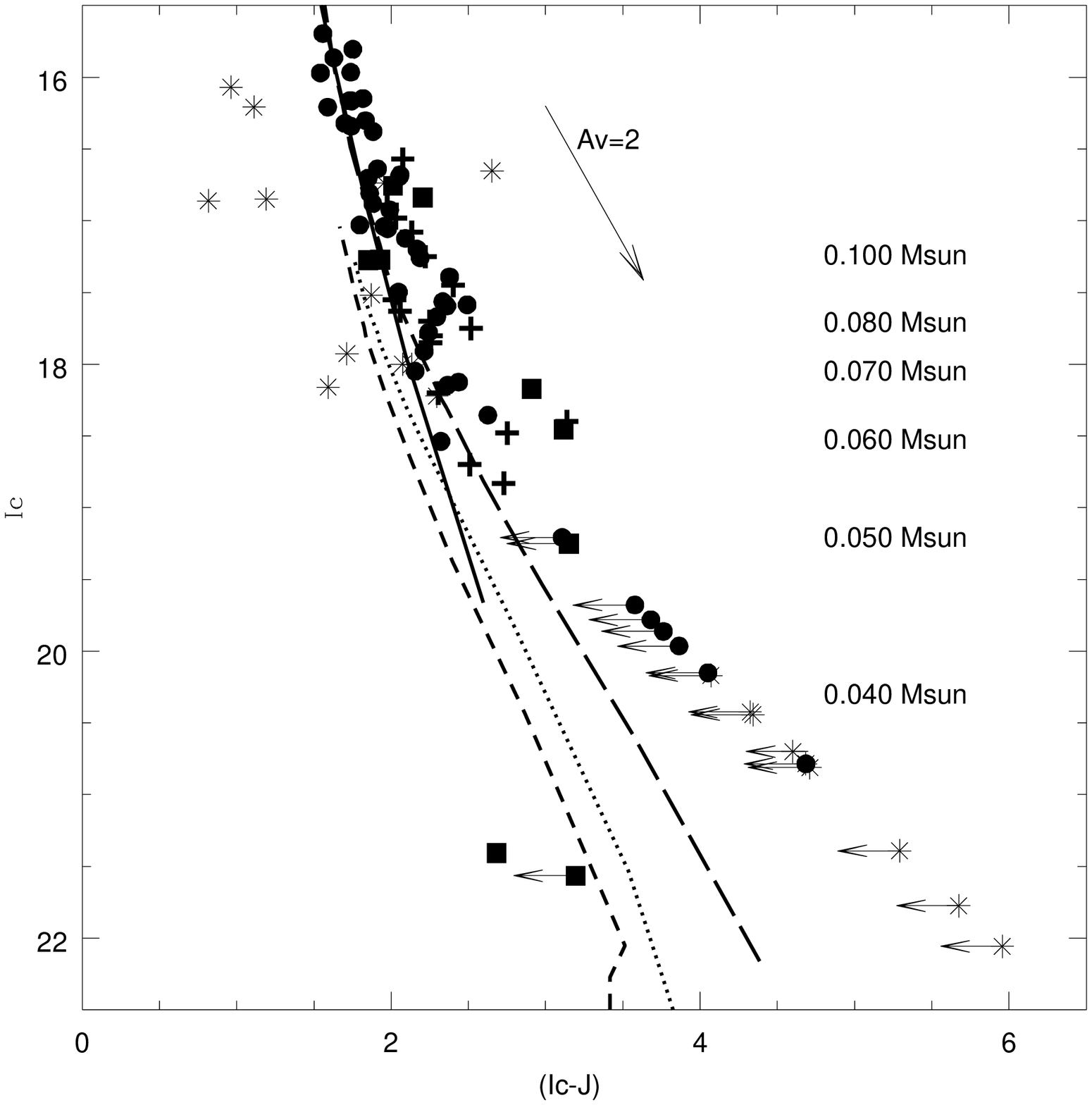}
      \caption{Colour--Magnitude diagrams of Alpha Per cluster.
See key for the meaning of the symbols.
Three 80 Myr  isochrones from Baraffe et al. (1998)  and 
Chabrier et al. (2000) (NextGen model as long-dashed lines,  
Dusty model as dotted lines,
and Cond model as short-dashed lines)
 are displayed together with 
a Leggett (1992) main sequence (solid lines).
For several $I$$_c$ magnitudes, the masses
 of cluster members are  indicated in 
the right-hand side of each 
panel. We have included a reddening vector for
 A$_V$=2 for comparison purposes.}
         \label{Fig3}
   \end{figure*}

\setcounter{figure}{2}

   \begin{figure*}
   \centering
   \includegraphics[width=8cm]{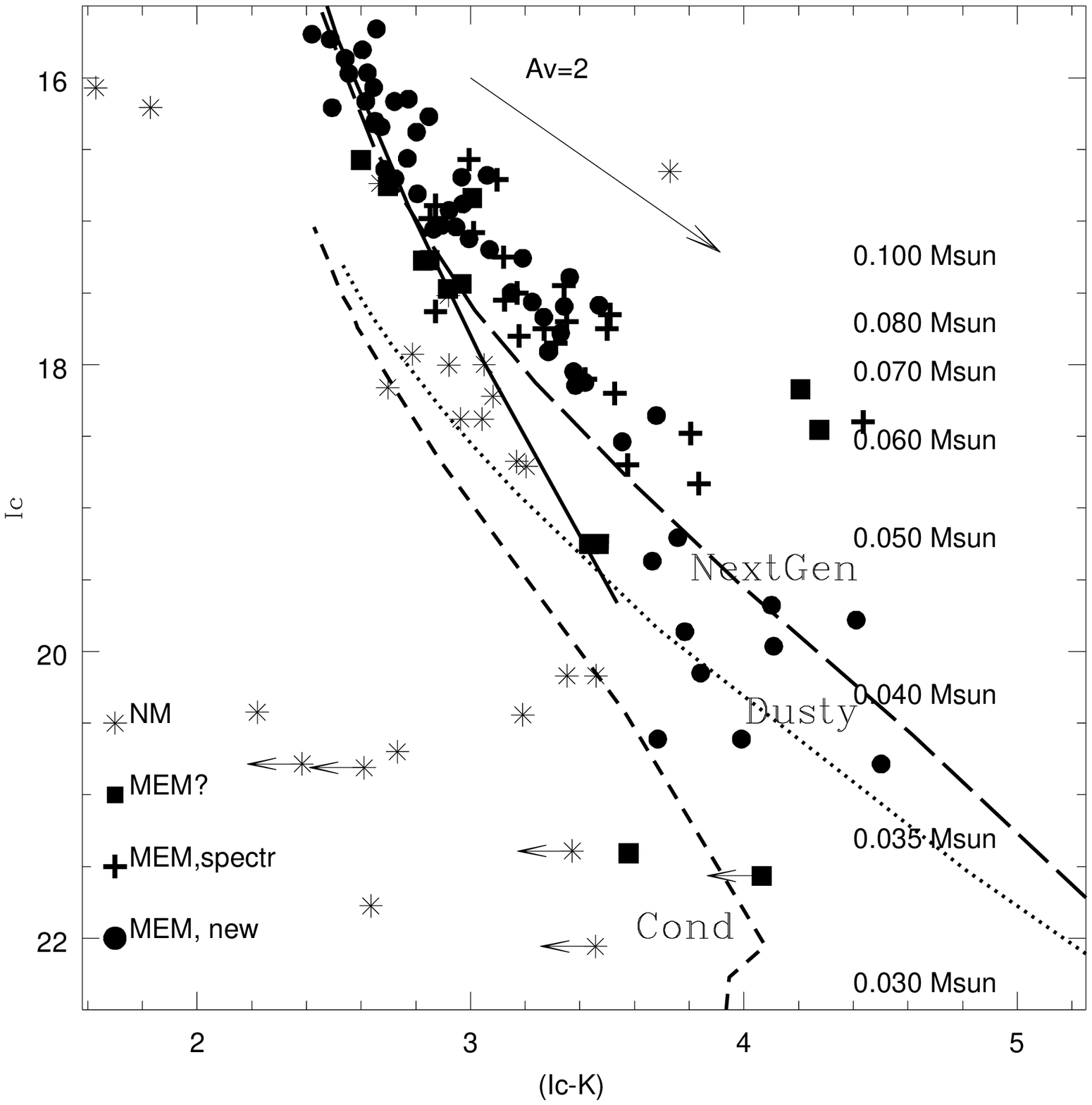}
   \includegraphics[width=8cm]{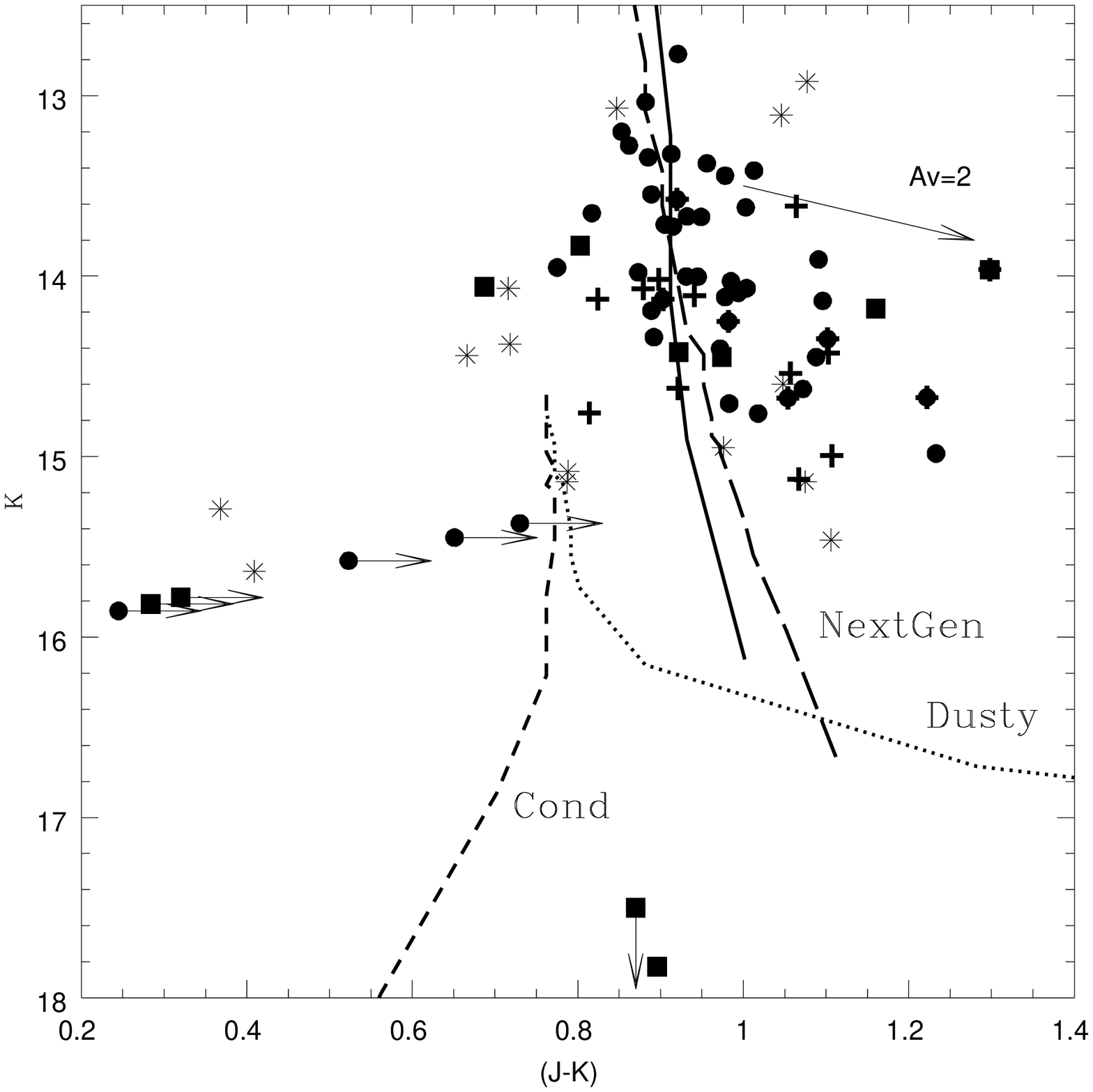}
      \caption{(continue figure 3)}
         \label{Fig3}
   \end{figure*}

We also searched the 2MASS point source catalogue (Skrutskie
et al. 1997), in order to extend the infrared data coverage. Unfortunately,
the present version of the 2MASS catalogue does not completely cover the
Alpha Per cluster, due to gaps between the scans in that part of the sky.
Moreover, the 2MASS survey does not go deep enough to detect all faint
candidates in our survey. In summary, 47 of our candidates were found
in the 2MASS catalogue; 25 additional 
candidates lie in the non-covered areas; and
21 are too faint to have been detected.

In total, all but two BD candidates derived from the optical
KPNO MOSA data also have near-infrared photometry. Table 3 lists all
available photometry.


\subsection{Previously known Alpha Per members}

In the following analysis of membership, we have also
included previously-discovered photometrically-selected members of the 
cluster (Stauffer et al. 1999). Some of these also have low- and 
medium-resolution spectroscopy which supports their membership.
We also searched the 2MASS catalogue for infrared photometry for these
previously-known members (Table 2), for comparison with our MOSA-selected 
sample. Note, of course, that Alpha Per cluster contains other brighter and more
massive members not analysed here (Stauffer et al. 1986, 1989;
Prosser 1992, 1994, 1998).


\section{Analysis}

\subsection{Optical photometry and coordinates}

Raw photometry was derived using PSF--fitting, and calibrated using 
several standard stars (see section 2.1). We computed the 
initial coordinates by transforming the X,Y positions of each CCD to
$\alpha$, $\delta$ using field
stars with known coordinates. The accuracy of this astrometry is better
than 1 arcsec. Following this, we searched for the 2MASS counterparts, and
when available, we used the much more accurate 2MASS astrometric positions
in Table 3.

Our KPNO/MOSA optical data cover the range 16$\le$$I$$_c$$\le$22.5,
0.0$\le$($R$-$I$)$_c$$\le$3.5. Figure 1 allows an
estimation of the  completeness limits. We have displayed two
histograms with the number of detections in the $R$$_c$ and $I$$_c$ filters
in a logarithmic form. The detections limits are located at the
 points where the histograms diverge from the straight lines
(Wainscoat et al. 1992; Santiago et al. 1996).
This is true for a region where
field stars dominate the overall population, as in the case of the 
Alpha Per cluster (many more field stars than cluster members in a
particular area). 
Therefore, for cluster members, we estimate that
$I_{complete}$$\sim$20.75 mag,
$R_{complete}$$\sim$21.9 mag.
However, since the $R$$_c$ filter effectively limits  the completeness
for cluster members, the actual values are
$I_{complete}$$\sim$19.5 mag,
$(R-I)_{complete}$$\sim$2.4.
The detection limits reach
$I_{limit}$$\sim$22.5 mag,
$R_{limit}$$\sim$25.5 mag.


\subsection{Optical colour-magnitude diagram}

Figure 2 displays all the detections 
 within our MOSA
survey area. The straight dashed line denotes
 the position of a fiducial 
main sequence (i.e., the criterion we have used to 
select candidate members of the cluster).
This line is  an empirical zero age main sequence,
shifted to fit  the locus of previously known Alpha Per members.
A total of 260,000 stars were detected, by far the majority
of them field stars, fairly well separated from the cluster population
by a fairly wide gap. This gap is not as sharp however as is
seen in the Pleiades (Figure 2 of Bouvier et al. 1998), and thus we
anticipate a stronger contamination of the cluster population in Alpha
Per compared to the Pleiades: the latter cluster has a contamination
of $\sim$30\% (Bouvier et al. 1998; Mart\'{\i}n et al. 2000; Moraux
et al. 2001).

We have selected 94 candidate members based on the optical CMD.
Since some of our fields overlap with each other, some of them have 
been detected twice, showing similar photometry.
The candidate members  are listed in Table 3.
Figure 3a displays their location in an optical  CMD. 
Bona fide members from Stauffer et al. (1999), based on 
optical-infrared data and spectroscopy are shown
as thick crosses. Thin crosses denote the location 
of new candidates with no infrared data. Non-members, based on the
new available data (see next section), appear as asterisks.
 Finally, possible and probable new members are
shown as solid squares and circles, respectively.
Also plotted are 
three 80 Myr  isochrones from Baraffe et al. (1998)  and 
Chabrier et al. (2000) (NextGen model as long-dashed line,  
Dusty model as dotted line,
and Cond model as short-dashed line), along with a
Leggett (1992) empirical main sequence for the cluster, plotted as
a solid  line.
The $I$$_c$ magnitudes  of Alpha Per members with 
different masses, derived with a 80 Myr NextGen isochone, are also indicated.

Note that all colour-magnitude and colour-colour diagrams
presented in this paper (this and next subsections) have made use of the 
following reddening values:
A$_I$=0.179, A$_J$=0.085, A$_K$=0.034, and
E($V$-$I$$_c$)=0.121, E($R$-$I$)$_c$=0.067, E($I$$_c$-$J$)=0.094, 
E($I$$_c$-$K$)=0.145, and E($J$-$K$)=0.051.
These values were derived from 
A$_V$=0.30, R$_V$=3.12 and  the interstellar extinction law and
transformation equation between filters and systems
published by Rieke \& Lebofsky (1985)
 and Taylor (1986).

%
%
   \begin{figure}
   \centering
   \includegraphics[width=8cm]{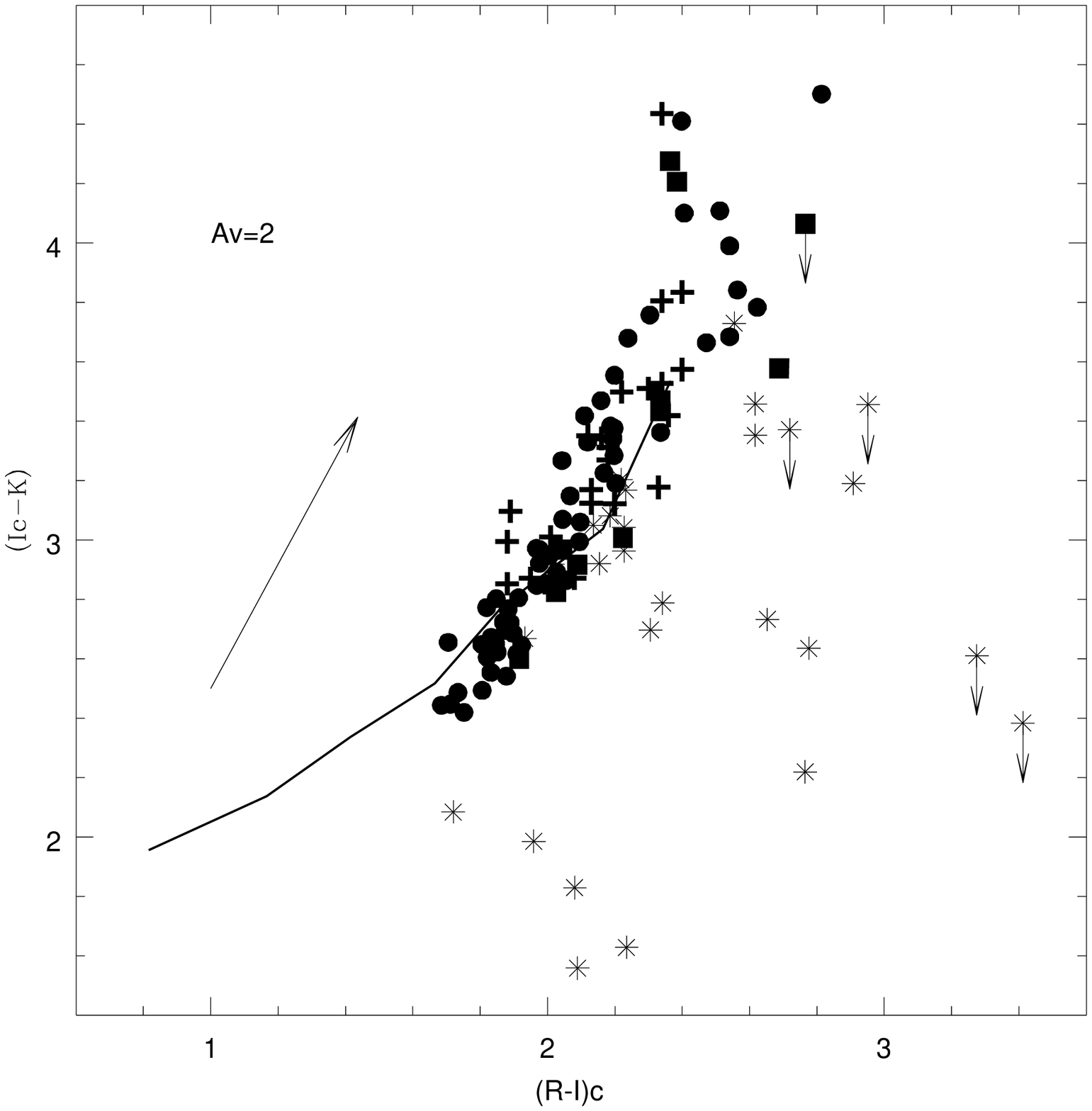}
      \caption{Optical-IR Colour-Colour Diagram for Alpha Per cluster.
Symbols as in Figure 3. The solid line comes from Leggett (1992).
}
         \label{Fig4}
   \end{figure}
%

\subsection{Infrared data and membership selection}

Alpha Per candidate members  brighter 
than I=18.5 are expected to be detected 
by the 2MASS survey (Skrutskie et al. 1997), which 
reaches about $K$$\sim$15.
We have searched in this database and extracted the 
IR photometry for them. However, 2MASS has several patches
in the Alpha Per area, with gaps between the images and some 
of our candidates were not observed.
For objects at the faint end of  2MASS and  fainter than it, 
we have carried out our own deep IR observations.
All these data are listed in Table 3.

Figures 3b, c, and d show optical/IR-colour magnitude
diagrams for the candidates derived from the KPNO, Calar Alto, and 2MASS
data (Table 3).
 Symbols are as in Figure 3a (see key or section 3.2).
As in the case of panel a, 
a reddening vector for A$_V$=2.0 has been included for comparison
purposes.
Note that neither of the two isochrones fully agree either
with the new MOSA data or with the previously discovered
bona fide members (Stauffer et al. 1999).
This is perhaps not unexpected, as even these models which include
 state-of-the-art model atmospheres do not as yet accurately reproduce
 the optical SED's for late type stars and hence, in particular, the
 R band magnitudes are only approximate for the very cool objects in
 our figures.
Figure 4 shows  an optical-infrared colour-colour diagram for 
the same sample. The solid line corresponds
to an empirical main sequence based on Leggett (1992) data.
Objects with full agreement between their photometric locations
in the four colour-magnitude diagrams and those expected for bona fide
cluster members are henceforth classified as probable cluster members.
Others, with slight disagreements in their photometry are classified
as possible members. Finally, some (shown as asterisks) could be readily
rejected as cluster members. The classification of each object is shown
in the last column  of Table 3.
Out of the initial 94 candidates, more than half (54) have
been classified as probably cluster members and 12 as possibles, while
26 objects turn out to be non-members. The remaining 2 candidates do
not have infrared photometry and cannot be classified. Thus, the 
contamination of the sample lies in the range 30--45\%, similar or larger 
than in the case of the Pleiades (Bouvier et al. 1998; Mart\'{\i}n et al. 2000;
Moraux et al. 2001).


\subsection{Luminosity and Mass Functions}

   \begin{figure}
   \centering
   \includegraphics[width=8cm]{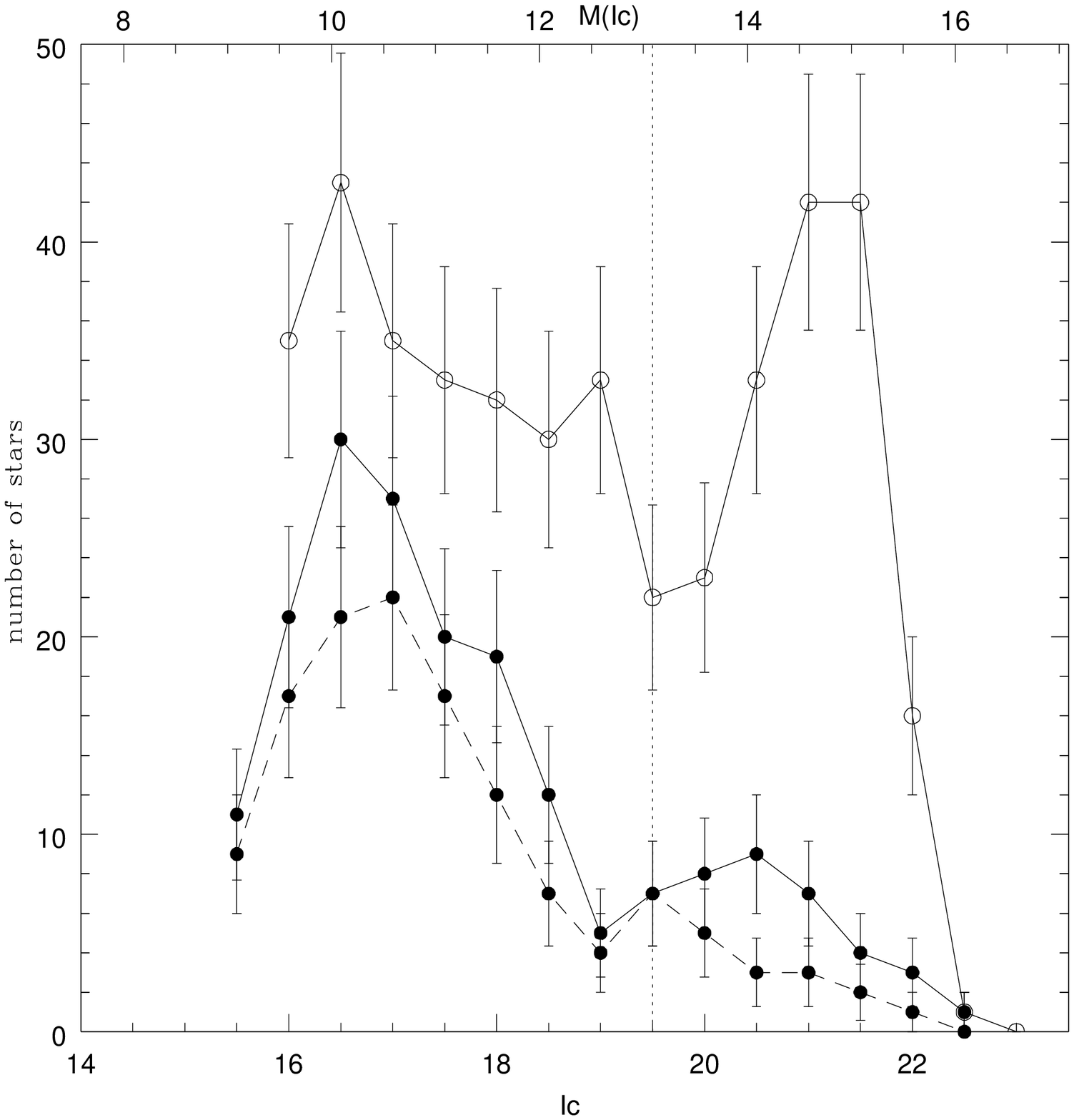}
      \caption{Luminosity Function of Alpha Per cluster.
The lower LFs correspond (solid circles) to the initial MOSA candidate members 
of the cluster,
 whereas the upper curve (open circles) was derived for a comparison
sample of field stars. 
In the first case, we show the LF before and after rejecting
the non-members (solid and dashed lines, respectively).
}
         \label{Fig5}
   \end{figure}
%

\subsubsection{Luminosity Function}

 We have computed the 
cluster luminosity function (LF) at the
end of the main sequence. We have done this both 
before and 
after rejecting the non-members from our initial
membership list. The LF is illustrated in Figure 5, where
we represent the histograms computed 
with the initial  and final candidate members as solid circles
(solid and dashed lines, respectively) and
a control group of objects selected in the same fashion,
 but with colours bluer than them, with an arbitrary shift of 
$\Delta$($R$-$I$)=--0.2 mag.
Based on this
selection criterion, the vast
majority of the comparison sample should be field stars, at a
distance similar to that of the cluster stars.
The dotted 
segment  represents the location of the completeness limit,
which corresponds to $\sim$0.05 M$_\odot$. Poisson noise
for each value is represented with the vertical error bar.
Note that apparent $I$$_c$ magnitudes are displayed at the bottom
x-axis, whereas absolute magnitudes are represented at the top x-axis.
The Alpha Per cluster LF presents two peaks (at this magnitude range): 
one at M(I$_c$)$\sim$10 and another at 14.5 magnitudes.
In the first case, the peak appears both in the Alpha Per 
member list and in the control group described above. In the second case,
the sample is not complete, and it could be strongly influenced
by pollution by spurious members, as the comparison with the control 
group suggests.
Note that the peak at M(I$_c$)$\sim$10 is
present in some other young clusters,
 including the Pleiades (Zapatero Osorio 1997), but
not all, for example, NGC\,2516 (Jeffries et al. 2001; 
Barrado y Navascu\'es et al. 2002),
 a rich, $\sim$125 Myr old
cluster that is either slightly
metal deficient (Jeffries et al. 1997, 1998, 2001) or approximately solar
metallicity (Debernardi and North 2001; Terndrup et al. 2002).
The peak in the LF for NGC2516 appears to occur about a magnitude
fainter than for the Pleiades and Alpha Per (that is,
at about M(I$_c$)$\sim$11).  Another similar age cluster,
M35, appears to have the peak occur at about
M(I$_c$)$\sim$9 (Barrado y Navascues et al 2001a).   Therefore,
the exact absolute magnitude of the peak
does not seem to be universal, at least
for 100-200 Myr clusters.

 The gap in the  Alpha Per LF at M(I$_c$)$\sim$12.5
(corresponding to M6-M8 spectral type or 
about 0.055 M$_\odot$ for an age of 80 Myr) has been also found
in other clusters. This is the case of the young cluster IC2391 (Barrado y
Navascu\'es et al. 2001b). Jameson (2002) noticed that this gap, which appears
at about M7 spectral type, is present in clusters with ages ranging from 
few million years (Sigma Orionis cluster), up to at least seven hundred
million year (Praesepe). It is clearly seen in the Pleiades too.
They suggested that it could be related to the 
formation at these temperatures of large  dust grains, although it might be related
to some specific behavior in the luminosity--mass relationship or
to the formation mechanism of these low mass objects
(the efficiency of the fragmentation and collapse for
cores with this mass range within the parental molecular cloud).

%
   \begin{figure}
   \centering
   \includegraphics[width=8cm]{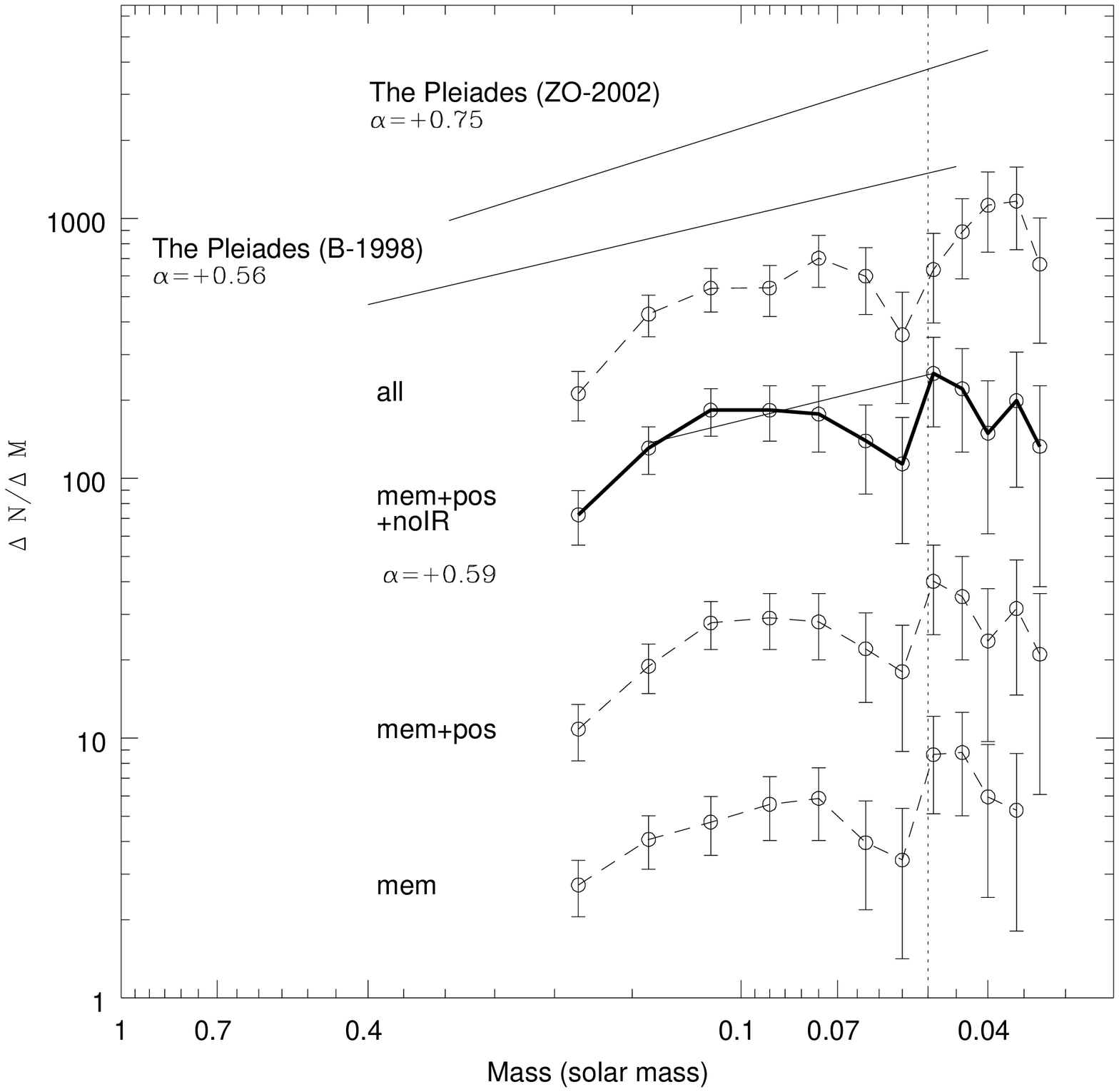}
      \caption{Mass functions for the
Alpha Persei cluster, illustrating the effect of contamination by
field stars.
These MFs were derived using  a 80 Myr isochrone (Baraffe et al. 1998).
The most probable MF is shown as a thick line.
A linear fit for this MF is also included, with  a power law index
of $\alpha$=0.59$\pm$0.05.
For comparison, we have included the Pleiades MF by Bouvier et al. (1998),
valid between 0.40 and 0.045 M$_\odot$ and by Zapatero Osorio el at. (2002),
which was derived for the mass range 0.3--0.040  M$_\odot$.
}
         \label{Fig6}
   \end{figure}
%

%
\subsubsection{Mass Function}

We have computed the Mass Function of the Alpha Per cluster
using the non-dusty (Next-Gen) models by Baraffe et al. (1998)
Note that the global behavior of the MF is not strongly affected
by the particular set of isochrones, as we have shown in
the case of M35 (Barrado y Navascu\'es et al. 2001a)
and NGC2516 (Barrado y Navascu\'es et al. 2002), using 
evolutionary  tracks from Baraffe et al. (1998),
D'Antona \& Mazzitelli (1994, 1997) and Siess et al. (2000).
In order to assess the effects of contamination by field
stars in our survey, and biases due to incomplete coverage in the infrared,
we have estimated the MF based on several different subsamples of the
candidate list. For the first sample, we used the full candidate list 
derived from the optical data only. In the second, we removed definite
non-members. In the third, we removed sources without infrared photometry.
Finally, in the fourth, we included only those sources classified as
probable cluster members. All four derived MFs are shown in Figure 6
along with the MF for the Pleiades (Bouvier et al. 1998; Moraux
et al. 2001; Zapatero Osorio et al. 2002).

Overall, the behaviour of the Alpha Per cluster MF
is very similar to that of the Pleiades, both in the overlapping 
mass range and for less massive brown dwarfs.
However, the MF of the Alpha Per cluster 
shows some possible  structure, with a dip about 0.055
M$_\odot$ and a drop beyond 0.035 M$_\odot$, regardless the 
cluster age in the range 50--100 Myr  (see next paragraph).
Unfortunately, the
uncertainties are large, the completeness limit of our MOSA survey
is about 0.050 M$_\odot$, and there is significant pollution by
field stars below this mass (peaking at about 0.030 M$_\odot$).

Figure 7 shows several realisations of the Alpha
Per MF as a function of age (50, 80, and 100 Myr), using the complete
list of candidates minus the definite non-members (Section 3.2). 
MFs 
have been derived for both the non-dusty Baraffe
et al. (1998) and Burrows et al.  (1997).
Masses were derived either from the M($I_c$) magnitude
(Baraffe et al. 1998 list these values) or from the 
bolometric luminosities (after Monet et al. 1992).
It is readily
 seen that the slope, within the uncertainties, 
 and the general shape of the cluster MF is
 essentially independent of the adopted age or the model.  The exact mass of
 the dip in the MF (assuming it is real), however, does shift
 slightly depending on the age.

When defining the MF as $\phi$$\propto$$M^{-\alpha}$ 
(or more properly, the mass spectrum), the spectral index $\alpha$
indicates the slope in the diagram we show in Figure 6.
We have fit a power law function to this MF, obtaining 
$\alpha$=0.59$\pm$0.05.
For the Pleiades, this index is 0.6, as derived by
Bouvier et al. (1998). See also Mart\'{\i}n et al. (2000);
 Hodgkin \& Jameson (2000); 
Moraux et al. (2001); Zapatero Osorio et al. (2002)
--$\alpha$=0.75;
 and references therein. 
Two of the Pleiades MFs are shown in Figure 6.
B\'ejar et al. (2001) have derived a MF for the  younger cluster
Sigma Orionis, about 5 Myr, down to the planetary mass 
domain at $\sim$0.013 M$_\odot$. This MF is shown in Figure 7. 
The spectral index of Sigma Orionis cluster
($\alpha$=0.8$\pm$0.4) also agrees quite well, within the uncertainties,
 with both Alpha Per cluster
and the Pleiades. 
Note that the Alpha Per  data corresponding to 
 the gap at 0.055  M$_\odot$ 
have not been included in this fit.
In any case, 
independently of the power law index of the MF,
 Figures 6 and 7 clearly show that the MF keeps
rising  well below the substellar limit for Alpha Per cluster.

%
   \begin{figure}
   \centering
   \includegraphics[width=8cm]{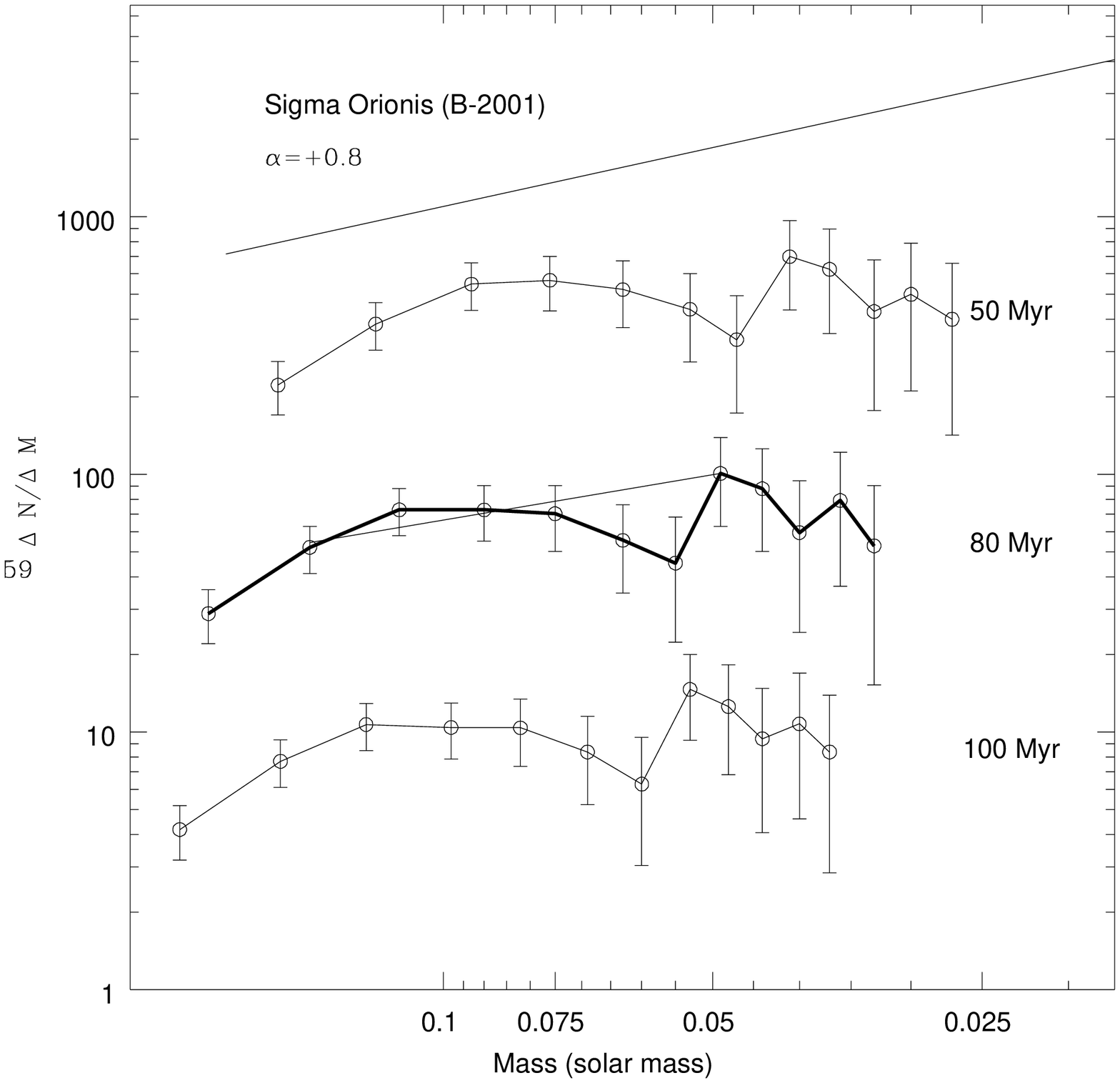}
   \includegraphics[width=8cm]{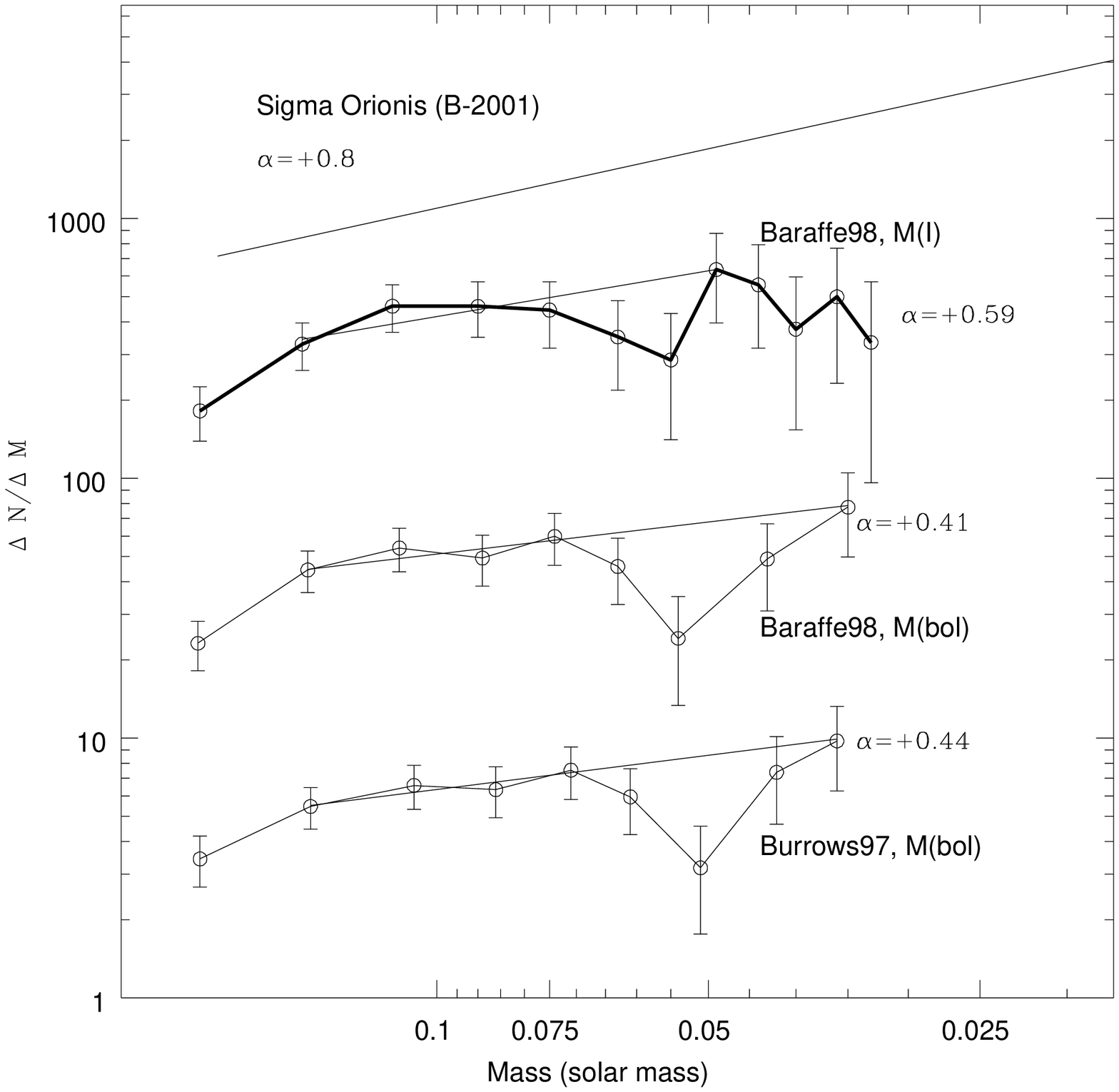}
      \caption{Mass functions of
Alpha Persei, illustrating the effect of the assumed age (panel a) and the
choice of theoretical models  (panel b).
A power law fit and its slope is included in each case.
 A MF for the Sigma Orionis cluster  (B\'ejar et al. 2001) 
is shown for comparison purposes (solid segment).
}
         \label{Fig7}
   \end{figure}
%


\subsection{Brown dwarfs in the cluster}

Recently, both Stauffer et al. (1999) and Basri \& Mart\'{\i}n (1999)
 searched for evidence of BDs in the Alpha Per cluster.
Previously, Rebolo et al. (1992) and Zapatero Osorio et al. (1996)
tried, without success, to find them
in this association.
Stauffer et al. (1999) included both low and medium  resolution 
spectroscopy, and some infrared photometry, establishing
the location of the lithium depletion boundary
(with the lithium age of the cluster, 90$\pm$10 Myr) and
the member list of bona fide members (based on rough radial
velocity, lithium detection, H$\alpha$ emission, spectral types, 
{\it et cetera}).  Since their
spectroscopic data go down to $I$$_c$=18.7, and the
stellar/substellar boundary is located at $I$$_c$=18.2
(Baraffe et al. 1998), they were able to catalogue
2 objects as probable brown dwarfs (another 3, which were
 catalogued as  possible members, could also be BDs).
In our new MOSA sample, there are 27 objects below the substellar
frontier.  Out of these 27 objects, 
11  have infrared photometry which indicates
that they are probable members of the cluster
and, therefore,  brown dwarfs.
Another 5 have been classified as possible members and could
be BDs if their membership is confirmed.
The minimum mass is 0.035 M$_\odot$.
Therefore, it seems likely that we have discovered a substantial
population of BDs in the Alpha Per cluster.
Note that it is difficult to use the proper motion method 
to establish Alpha Per membership, since its proper motion
(22.93  mas/yr in $\alpha$ and --25.56 mas/yr in $\delta$,
 Robichon et al. 1999), is large enough to be  measured,
but is also shared by a large fraction of the field stars (see, e.g.,
Prosser 1992). Therefore, 
only additional spectroscopic data 
(or better, a combination of optical and infrared photometry, 
proper motions and spectroscopy) will demonstrate beyond
any reasonable doubt that the faint objects listed in our sample
are, indeed, brown dwarfs.


\section{Summary}

Optical and near infrared photometry have been used to
select a list of probable and possible low-mass  members 
belonging to  the
young cluster Alpha Per, unveiling a large population
of brown dwarf candidates.
This information has been used to derive the luminosity and mass function
of the cluster in the substellar domain.
The index of the mass function, $\alpha = 0.59$,
is very similar to that determined for the Pleiades  (120 Myr) and
$\sigma$ Orionis (2--8 Myr) clusters.


\begin{acknowledgements}
DBN and JRS acknowledge the hospitality 
of the Observatoire de Grenoble 
during the preparation of this paper. 
We thank D. Montes for his help collecting part of the IR data
and M.R. Zapatero Osorio for her comments
 on an early version of this paper.
JB, NL, and MJM acknowledge support from the 
EC Research Training Network ``The
Formation and Evolution of Young Stellar Clusters'' (HPRN-CT-2000-00155).
This work has been partially financed by 
Spanish {\it ``Programa Ram\'on y Cajal''} 
and AYA2001-1124-CO2 programs.
\end{acknowledgements}

\setcounter{table}{1}
\begin{table}
\caption[]{Previusly known Alpha Per members from
Stauffer et al. (1999). Data from 2MASS survey.}
\begin{tabular}{lccc}
\hline
  Name      &     J           &      H          &    Ks           \\
\hline
 AP300      &   15.595  0.061 &   14.859  0.074 &   14.538  0.083 \\ 
 AP301      &   15.233  0.061 &   14.526  0.072 &   14.251  0.082 \\
 AP302      &   15.572  0.066 &   15.101  0.092 &   14.758  0.102 \\
 AP303      &   14.951  0.043 &   14.278  0.049 &   14.127  0.062 \\
 AP304      &   16.102  0.107 &   15.418  0.134 &   14.995  0.130 \\
 AP305      &   15.729  0.077 &   14.906  0.084 &   14.675  0.100 \\
 AP306      &   16.240  0.111 &   15.727  0.160 &   15.077   null \\
 AP306      &   15.262  0.050 &   14.522  0.063 &   13.964  0.051 \\ 
 AP307      &   14.948  0.043 &   14.280  0.055 &   14.069  0.060 \\ 
 AP308      &   14.677  0.035 &   14.128  0.048 &   13.613  0.049 \\ 
 AP309      &   14.495  0.032 &   13.882  0.044 &   13.575  0.045 \\ 
 AP275      &   15.031  0.046 &   14.488  0.059 &   14.128  0.066 \\ 
 AP310      &   15.543  0.065 &   15.110  0.104 &   14.622  0.085 \\ 
 AP311      &   15.450  0.058 &   14.709  0.069 &   14.348  0.069 \\ 
 AP312      &   16.274  0.118 &   15.483  0.131 &   15.132  0.135 \\ 
 AP313      &   15.528  0.066 &   15.035  0.103 &   14.425  0.081 \\ 
 AP314      &   16.196  0.107 &   15.433  0.128 &   15.133  0.151 \\ 
 AP315      &   15.894  0.089 &   15.234  0.111 &   14.672  0.104 \\ 
 AP318      &   15.049  0.049 &   14.502  0.068 &   14.108  0.071 \\ 
 AP319      &   14.916  0.053 &   14.378  0.063 &   14.018  0.069 \\ 
 AP326      &   16.192  0.108 &   15.556  0.125 &   15.125  0.149 \\ 
\hline
\end{tabular}
$\,$
\end{table}

\setcounter{table}{2}
\begin{table*}
\caption[]{\tiny $\alpha$ Per data from KPNO/MOSA 1998. Probable members, possible members, probable non-members, and no IR data.}
{\tiny
\begin{tabular}{lc cccc ccccccc}
 ID      &  alpha        delta     & I$_c$  &(R-I)$_c$&$\delta$I $\delta$R& J         &    H            &   Ks           &  J             &     K'         & Source  & Mem. \\
\cline{6-8}
         &     (J2000.0)           &        &       &                   &         \multicolumn{3}{c}{2MASS}              &                &                &         &        \\
\hline
\hline
AP282   & 3 24 22.55  +48 24 26.1 & 15.217 & 1.712 & 0.020  0.002 &   13.690 0.030  &   13.019 0.035  &   12.769 0.033 &    --     --   &    --     --   & --      & Y   \\  
AP283   & 3 24 38.78  +48 17 17.1 & 15.696 & 1.752 & 0.015  0.003 &   14.138 0.030  &   13.520 0.036  &   13.276 0.044 &    --     --   &    --     --   & --      & Y   \\  
AP301   & 3 18 09.05  +49 25 19.0 & 17.594 & 2.194 & 0.004  0.005 &   15.233 0.062  &   14.526 0.072  &   14.251 0.082 &   14.970 0.019 &   13.854 0.021 & b       & Y   \\  
AP305   & 3 19 21.60  +49 23 31.0 & 18.355 & 2.239 & 0.009  0.012 &   15.729 0.078  &   14.906 0.084  &   14.675 0.100 &   15.774 0.014 &   14.703 0.018 & b       & Y   \\  
AP306   & 3 19 41.32  +50 30 45.1 & 18.171 & 2.385 & 0.003  0.008 &   15.262 0.051  &   14.522 0.063  &   13.964 0.051 &    --     --   &    --     --   & --      & Y?  \\  
AP309   & 3 22 40.65  +48 00 33.6 & 16.378 & 1.848 & 0.003  0.002 &   14.495 0.035  &   13.882 0.045  &   13.575 0.046 &    --     --   &    --     --   & --      & Y   \\  
AP311   & 3 23 08.67  +48 04 50.6 & 17.497 & 2.067 & 0.004  0.005 &   15.450 0.060  &   14.709 0.070  &   14.348 0.070 &    --     --   &    --     --   & --      & Y   \\  
AP315   & 3 26 34.10  +49 07 46.1 & 18.049 & 2.198 & 0.009  0.012 &   15.894 0.090  &   15.234 0.111  &   14.672 0.104 &   15.899 0.022 &   14.982 0.032 & b       & Y   \\  
AP327   & 3 20 31.74  +49 39 59.6 & 15.056 & 1.959 & 0.089  0.016 &   13.917 0.035  &   13.388 0.038  &   13.070 0.037 &    --     --   &    --     --   & --      & N   \\  
AP328   & 3 21 25.61  +48 10 01.0 & 15.193 & 1.720 & 0.083  0.015 &   14.154 0.033  &   13.397 0.037  &   13.108 0.038 &    --     --   &    --     --   & --      & N   \\  
AP329   & 3 23 56.34  +48 09 21.0 & 15.479 & 1.685 & 0.026  0.002 &   13.917 0.030  &   13.290 0.033  &   13.035 0.035 &    --     --   &    --     --   & --      & Y   \\  
AP330   & 3 34 15.6   +49 58 48   & 15.659 & 1.705 & 0.006  0.027 &     --    --    &     --    --    &     --    --   &    --     --   &   13.003 0.008 & d       & Y   \\  
AP331   & 3 32 05.9   +50 05 55   & 15.731 & 1.734 & 0.033  0.042 &     --    --    &     --    --    &     --    --   &    --     --   &   13.244 0.009 & d       & Y   \\  
AP332   & 3 25 16.90  +48 36 09.0 & 15.805 & 1.821 & 0.003  0.010 &   14.053 0.036  &   13.410 0.040  &   13.200 0.035 &    --     --   &    --     --   & --      & Y   \\  
AP333   & 3 25 13.55  +50 27 33.0 & 15.865 & 1.878 & 0.014  0.008 &   14.236 0.031  &   13.603 0.034  &   13.323 0.039 &    --     --   &    --     --   & --      & Y   \\  
AP334   & 3 22 45.47  +48 21 33.1 & 15.965 & 1.850 & 0.013  0.004 &   14.227 0.034  &   13.641 0.041  &   13.342 0.036 &    --     --   &    --     --   & --      & Y   \\  
AP335   & 3 28 52.96  +50 19 25.9 & 15.970 & 1.832 & 0.034  0.008 &   14.428 0.041  &   13.761 0.040  &   13.415 0.047 &    --     --   &    --     --   & --      & Y   \\  
AP336   & 3 31 55.3   +49 08 31   & 16.039 & 1.846 & 0.026  0.214 &     --    --    &     --    --    &     --    --   &    --     --   &   14.659 0.031 & d       & N   \\  
AP337   & 3 34 07.5   +48 32 08   & 16.067 & 1.923 & 0.040  0.004 &     --    --    &     --    --    &     --    --   &    --     --   &   13.421 0.010 & d       & Y   \\  
AP338   & 3 24 52.65  +48 46 12.8 & 16.069 & 2.235 & 0.042  0.104 &   15.106 0.048  &   14.481 0.060  &   14.440 0.082 &    --     --   &    --     --   & --      & N   \\  
AP339   & 3 26 33.24  +50 07 41.7 & 16.148 & 1.819 & 0.043  0.003 &   14.331 0.037  &   13.734 0.037  &   13.375 0.037 &    --     --   &    --     --   & --      & Y   \\  
AP340   & 3 21 34.85  +48 16 28.7 & 16.164 & 1.909 & 0.002  0.002 &   14.436 0.034  &   13.781 0.041  &   13.547 0.043 &    --     --   &    --     --   & --      & Y   \\  
AP341   & 3 31 03.39  +50 24 41.6 & 16.165 & 1.889 & 0.008  0.002 &   14.421 0.040  &   13.794 0.046  &   13.443 0.047 &    --     --   &    --     --   & --      & Y   \\  
AP342   & 3 25 39.24  +48 45 21.1 & 16.207 & 2.081 & 0.011  0.068 &   15.095 0.044  &   14.447 0.057  &   14.377 0.075 &    --     --   &    --     --   & --      & N   \\  
AP343   & 3 23 48.48  +48 36 43.0 & 16.208 & 1.806 & 0.007  0.012 &   14.619 0.045  &   14.003 0.050  &   13.714 0.052 &    --     --   &    --     --   & --      & Y   \\  
AP344   & 3 26 52.0   +50 00 33   & 16.270 & 1.968 & 0.022  0.002 &     --    --    &     --    --    &     --    --   &    --     --   &   13.422 0.006 & c       & Y   \\  
AP345   & 3 33 45.8   +50 08 53   & 16.302 & 1.846 & 0.005  0.008 &     --    --    &     --    --    &     --    --   &   14.468 0.010 &   13.651 0.012 & d       & Y   \\  
AP346   & 3 21 30.06  +48 49 23.2 & 16.321 & 1.805 & 0.050  0.022 &   14.621 0.037  &   13.979 0.045  &   13.672 0.047 &    --     --   &    --     --   & --      & Y   \\  
AP347   & 3 31 33.77  +49 52 02.1 & 16.342 & 1.833 & 0.007  0.004 &   14.601 0.045  &   13.967 0.050  &   13.669 0.055 &    --     --   &    --     --   & --      & Y   \\  
AP348   & 3 26 48.2   +48 44 00   & 16.466 & 1.926 & 0.032  0.009 &     --    --    &     --    --    &     --    --   &    --     --   &    --     --   & --      & --  \\  
AP349   & 3 26 47.9   +50 02 16   & 16.562 & 1.883 & 0.011  0.002 &     --    --    &     --    --    &     --    --   &    --     --   &   13.793 0.005 & c       & Y   \\  
AP350   & 3 32 06.8   +49 25 23   & 16.574 & 1.916 & 0.024  0.009 &     --    --    &     --    --    &     --    --   &    --     --   &   13.974 0.020 & d       & Y?  \\  
AP351   & 3 28 47.8   +50 02 01   & 16.638 & 1.898 & 0.008  0.004 &     --    --    &     --    --    &     --    --   &   14.727 0.033 &   13.952 0.019 & d       & Y   \\  
AP352   & 3 19 51.98  +48 48 22.6 & 16.652 & 2.555 & 0.009  0.007 &   13.999 0.030  &   13.344 0.035  &   12.922 0.037 &    --     --   &    --     --   & --      & N   \\  
AP353   & 3 24 48.66  +48 49 47.0 & 16.680 & 2.097 & 0.004  0.005 &   14.622 0.037  &   13.975 0.049  &   13.619 0.051 &    --     --   &    --     --   & --      & Y   \\  
AP354   & 3 27 31.64  +48 53 23.3 & 16.693 & 1.976 & 0.009  0.024 &   14.640 0.043  &   14.104 0.052  &   13.725 0.056 &    --     --   &    --     --   & --      & Y   \\  
AP355   & 3 22 33.15  +48 47 00.3 & 16.703 & 1.869 & 0.009  0.005 &   14.852 0.043  &   14.255 0.059  &   13.979 0.058 &    --     --   &    --     --   & --      & Y   \\  
AP356   & 3 18 22.06  +49 20 19.4 & 16.737 & 1.933 & 0.009  0.004 &   14.784 0.047  &   14.240 0.064  &   14.068 0.064 &    --     --   &    --     --   & --      & N   \\  
AP357   & 3 32 42.5   +49 50 10   & 16.740 & 2.068 & 0.015  0.004 &     --    --    &     --    --    &     --    --   &    --     --   &    --     --   & --      & --  \\  
AP357   & ``           ``         & 16.748 & 2.090 & 0.002  0.004 &     --    --    &     --    --    &     --    --   &    --     --   &    --     --   & --      & --  \\  
AP358   & 3 34 37.8   +49 13 53   & 16.756 & 1.878 & 0.003  0.005 &     --    --    &     --    --    &     --    --   &   14.745 0.011 &   14.058 0.021 & d       & Y?  \\  
AP359   & 3 23 37.13  +48 37 15.3 & 16.809 & 1.914 & 0.002  0.004 &   14.948 0.045  &   14.296 0.057  &   14.003 0.057 &    --     --   &    --     --   & --      & Y   \\  
AP360   & 3 32 24.0   +50 16 58   & 16.838 & 2.224 & 0.002  0.013 &     --    --    &     --    --    &     --    --   &   14.634 0.010 &   13.831 0.016 & d       & Y?  \\  
AP361   & 3 23 14.03  +48 42 21.3 & 16.849 & 2.089 & 0.020  0.327 &   15.658 0.067  &   15.272 0.106  &   15.290 0.175 &    --     --   &    --     --   & --      & N   \\  
AP362   & 3 25 24.45  +48 45 21.3 & 16.862 & 1.902 & 0.008  0.070 &   16.045 0.087  &   15.800 0.161  &   15.636 0.225 &    --     --   &    --     --   & --      & N   \\  
AP363   & 3 24 00.33  +47 55 29.7 & 16.880 & 1.967 & 0.002  0.002 &   14.999 0.046  &   14.295 0.051  &   13.908 0.058 &    --     --   &    --     --   & --      & Y   \\  
AP364   & 3 20 39.16  +49 32 06.0 & 16.924 & 1.976 & 0.007  0.004 &   14.933 0.052  &   14.428 0.074  &   14.002 0.059 &    --     --   &    --     --   & --      & Y   \\  
AP365   & 3 28 22.94  +49 11 24.0 & 17.030 & 2.027 & 0.010  0.007 &   15.233 0.057  &   14.473 0.061  &   14.137 0.075 &    --     --   &    --     --   & --      & Y   \\  
AP366   & 3 26 35.50  +49 15 43.8 & 17.040 & 2.005 & 0.053  0.011 &   15.087 0.048  &   14.486 0.058  &   14.093 0.061 &    --     --   &    --     --   & --      & Y   \\  
AP367   & 3 34 40.80  +50 03 43   & 17.056 & 2.057 & 0.002  0.004 &    --     --    &     --    --    &     --    --   &   15.080 0.015 &   14.191 0.017 & d       & Y   \\  
AP368   & 3 23 03.38  +48 53 05.8 & 17.123 & 2.095 & 0.004  0.007 &   15.031 0.048  &   14.488 0.060  &   14.128 0.067 &    --     --   &    --     --   & --      & Y   \\  
AP368   & ``           ``         & 17.198 & 2.045 & 0.010  0.011 &   ``     ``     &   ``     ``     &   ``     ``    &    --     --   &    --     --   & --      & Y   \\  
AP369   & 3 26 44.9   +50 25 09   & 17.258 & 2.203 & 0.005  0.008 &    --     --    &     --    --    &     --    --   &   15.071 0.011 &   14.067 0.012 & d       & Y   \\  
AP370   & 3 24 21.28  +48 58 59.5 & 17.271 & 2.015 & 0.003  0.007 &   15.342 0.056  &   14.724 0.080  &   14.420 0.081 &    --     --   &    --     --   & --      & Y?  \\  
AP371   & 3 23 59.92  +48 54 10.7 & 17.274 & 2.026 & 0.002  0.009 &   15.422 0.063  &   14.724 0.077  &   14.448 0.086 &    --     --   &    --     --   & --      & Y?  \\  
AP372   & 3 28 02.20  +48 41 07.2 & 17.391 & 2.336 & 0.010  0.010 &   15.013 0.050  &   14.465 0.054  &   14.028 0.058 &    --     --   &    --     --   & --      & Y   \\  
AP373   & 3 33 20.6   +48 45 49   & 17.437 & 2.044 & 0.007  0.016 &     --    --    &     --    --    &     --    --   &    --     --   &   14.469 0.021 & d       & Y?  \\  
AP374   & 3 32 18.8   +49 32 18   & 17.472 & 2.089 & 0.004  0.006 &     --    --    &     --    --    &     --    --   &    --     --   &   14.554 0.022 & d       & Y?  \\  
AP375   & 3 20 43.49  +50 59 39.6 & 17.518 & 2.016 & 0.007  0.014 &   15.647 0.069  &   15.223 0.121  &   14.599 0.095 &   15.225 0.025 &   14.263 0.031 & b       & N   \\  
AP376   & 3 32 43.7   +50 18 25   & 17.564 & 2.168 & 0.005  0.015 &    --     --    &     --    --    &     --    --   &   15.230 0.019 &   14.338 0.029 & b       & Y   \\  
AP377   & 3 27 31.29  +48 39 23.3 & 17.586 & 2.159 & 0.015  0.019 &   15.094 0.055  &   14.482 0.071  &   14.116 0.071 &   14.896 0.007 &   13.905 0.018 & b       & Y   \\  
AP378   & 3 27 00.3   +49 14 38   & 17.670 & 2.043 & 0.010  0.010 &    --     --    &     --    --    &     --    --   &   15.374 0.025 &   14.402 0.032 & b       & Y   \\  
AP379   & 3 29 18.74  +50 22 10.9 & 17.779 & 2.119 & 0.005  0.009 &   15.537 0.071  &   15.085 0.088  &   14.449 0.079 &   15.456 0.025 &   14.672 0.040 & b       & Y   \\  
\hline
\hline
\end{tabular}
}
$\,$\\  
a.- 1999 November. b 2000 February. c 2000 December. d 2001 November. e 2001 December \\
H data from CAHA:
AP345 H=13.920$\pm$0.008, AP351 H=14.203$\pm$0.043, AP358 H=14.205$\pm$0.014, 
AP360 H=14.084$\pm$0.011, AP367 H=14.448$\pm$0.013, AP369 H=14.324$\pm$0.010\\
\end{table*}

\setcounter{table}{2}
\begin{table*}
\caption[]{\tiny (Continue) $\alpha$ Per data from KPNO/MOSA 1998. Probable members, possible members, probable non-members, and no IR data.}
{\tiny
\begin{tabular}{lc cccc ccccccc}
 ID      &  alpha        delta     & I$_c$  &(R-I)$_c$&$\delta$I $\delta$R& J         &    H            &   Ks           &  J             &     K'         & Source  & Mem. \\
\cline{6-8}
         &     (J2000.0)           &        &       &                   &         \multicolumn{3}{c}{2MASS}              &                &                &         &        \\
\hline
\hline
AP380   & 3 25 03.79  +48 49 58.6 & 17.910 & 2.198 & 0.004  0.008 &   15.697 0.072  &   15.078 0.100  &   14.625 0.102 &   15.374 0.025 &   14.355 0.030 & b       & Y   \\  
AP381   & 3 28 06.3   +48 45 13   & 17.927 & 2.342 & 0.009  0.013 &$>$16.1    --    &$>$15.6    --    &$>$15.3    --   &   16.214 0.045 &   15.139 0.062 & b       & N   \\  
AP382   & 3 33 48.08  +48 52 28   & 18.000 & 2.136 & 0.005  0.008 &     --    --    &     --    --    &     --    --   &   15.926 0.029 &   14.950 0.030 & b       & N   \\  
AP383   & 3 32 22.8   +48 54 11   & 18.003 & 2.154 & 0.003  0.005 &     --    --    &     --    --    &     --    --   &   15.870 0.023 &   15.082 0.039 & b       & N   \\  
AP384   & 3 27 40.03  +48 33 55.8 & 18.125 & 2.110 & 0.021  0.031 &   15.689 0.081  &   15.159 0.118  &   14.706 0.102 &   15.793 0.026 &   14.820 0.028 & b       & Y   \\  
AP385   & 3 20 07.42  +50 39 53.3 & 18.145 & 2.187 & 0.004  0.007 &   15.779 0.085  &   15.128 0.108  &   14.761 0.114 &   15.954 0.025 &   15.003 0.031 & b       & Y   \\  
AP386   & 3 22 02.3   +47 58 38   & 18.160 & 2.306 & 0.003  0.008 &$>$16.1    --    &$>$15.6    --    &$>$15.3    --   &   16.568 0.046 &   15.462 0.053 & b       & N   \\  
AP387   & 3 18 19.32  +49 14 35.1 & 18.222 & 2.186 & 0.004  0.007 &   15.927 0.106  &   15.293 0.129  &   15.140 0.162 &   15.955 0.022 &   15.015 0.031 & b       & N   \\  
AP388   & 3 31 47.4   +50 03 25   & 18.380 & 2.227 & 0.005  0.008 &     --    --    &     --    --    &     --    --   &    --     --   &   15.337 0.040 & d       & N   \\  
AP388   & ``           ``         & ``     & ``    & ``     ``    &     --    --    &     --    --    &     --    --   &    --     --   &   15.416 0.084 & e       & N   \\  
AP389   & 3 27 41.21  +48 40 33.9 & 18.455 & 2.364 & 0.019  0.022 &   15.339 0.067  &   14.647 0.073  &   14.179 0.074 &    --     --   &    --     --   & --      & Y?  \\  
AP390   & 3 21 54.45  +48 33 14.9 & 18.538 & 2.199 & 0.006  0.009 &   16.216 0.295  &   15.531 0.151  &   14.983 0.172 &    --     --   &    --     --   & --      & Y   \\  
AP391   & 3 32 48.1   +49 52 26   & 18.675 & 2.232 & 0.012  0.014 &     --    --    &     --    --    &     --    --   &    --     --   &   15.506 0.050 & d       & N   \\  
AP391   & ``           ``         & 18.710 & 2.218 & 0.005  0.008 &     --    --    &     --    --    &     --    --   &    --     --   &   ``     ``    & ``      & N   \\  
AP392   & 3 28 27.1   +50 20 02   & 19.207 & 2.304 & 0.006  0.014 &$>$16.1    --    &$>$15.6    --    &$>$15.3    --   &    --     --   &   15.449 0.017 & a       & Y   \\  
AP393   & 3 29 08.2   +48 25 35   & 19.250 & 2.335 & 0.022  0.035 &$>$16.1    --    &$>$15.6    --    &$>$15.3    --   &    --     --   &   15.780 0.021 & e       & Y?  \\  
AP393   & ``           ``         & ``     & ``    & ``     ``    &   ``      --    &   ``      --    &   ``      --   &    --     --   &   15.816 0.080 & d       & Y?  \\  
AP394   & 3 35 26.0   +49 46 55   & 19.370 & 2.472 & 0.008  0.017 &     --    --    &     --    --    &     --    --   &    --     --   &   15.705 0.023 & a       & Y   \\  
AP395   & 3 24 44.5   +48 21 29   & 19.678 & 2.406 & 0.024  0.036 &$>$16.1    --    &$>$15.6    --    &$>$15.3    --   &    --     --   &   15.577 0.023 & a       & Y   \\  
AP396   & 3 28 16.0   +48 37 40   & 19.781 & 2.398 & 0.016  0.048 &$>$16.1    --    &$>$15.6    --    &$>$15.3    --   &    --     --   &   15.370 0.015 & a       & Y   \\  
AP397   & 3 21 21.8   +48 02 23   & 19.862 & 2.623 & 0.012  0.038 &$>$16.1    --    &$>$15.6    --    &$>$15.3    --   &    --     --   &   16.078 0.029 & a       & Y   \\  
AP398   & 3 22 38.0   +48 58 03   & 19.964 & 2.512 & 0.011  0.028 &$>$16.1    --    &$>$15.6    --    &$>$15.3    --   &    --     --   &   15.855 0.023 & a       & Y   \\  
AP399   & 3 25 48.5   +50 00 59   & 20.151 & 2.564 & 0.014  0.045 &$>$16.1    --    &$>$15.6    --    &$>$15.3    --   &    --     --   &   16.309 0.047 & a       & Y   \\  
AP400   & 3 28 48.0   +50 20 42   & 20.170 & 2.617 & 0.019  0.085 &$>$16.1    --    &$>$15.6    --    &$>$15.3    --   &    --     --   &   16.711 0.175 & d       & N   \\  
AP400   & ``           ``         & ``     & ``    & ``     ``    &   ``      --    &   ``      --    &   ``      --   &    --     --   &   16.817 0.090 & e       & N   \\  
AP401   & 3 23 43.5   +48 22 39   & 20.424 & 2.765 & 0.020  0.292 &$>$16.1    --    &$>$15.6    --    &$>$15.3    --   &    --     --   &   18.204 0.140 & a       & N   \\  
AP402   & 3 30 38.7   +50 16 39   & 20.442 & 2.909 & 0.020  0.257 &$>$16.1    --    &$>$15.6    --    &$>$15.3    --   &    --     --   &   17.251 0.064 & e       & N   \\  
AP403   & ``           ``         & ``     & ``    & ``     ``    &     --    --    &     --    --    &     --    --   &    --     --   &   16.926 0.185 & d       & Y   \\  
AP403   & 3 32 04.6   +49 20 48   & 20.611 & 2.541 & 0.020  0.053 &     --    --    &     --    --    &     --    --   &    --     --   &   16.62  0.08  & e       & Y   \\  
AP404   & 3 23 48.3   +48 29 03   & 20.700 & 2.653 & 0.030  0.125 &$>$16.1    --    &$>$15.6    --    &$>$15.3    --   &    --     --   &   17.967 0.112 & a       & N   \\  
AP405   & 3 29 13.0   +50 05 24   & 20.784 & 3.412 & 0.064  0.224 &$>$16.1    --    &$>$15.6    --    &$>$15.3    --   &    --     --   &$>$18.4    --   & e       & N   \\  
AP406   & 3 23 09.9   +48 16 30   & 20.785 & 2.814 & 0.025  0.127 &$>$16.1    --    &$>$15.6    --    &$>$15.3    --   &    --     --   &   16.283 0.028 & a       & Y   \\  
AP407   & 3 23 59.3   +48 28 30   & 20.811 & 3.275 & 0.038  0.258 &$>$16.1    --    &$>$15.6    --    &$>$15.3    --   &    --     --   &$>$18.20   --   & e       & N   \\  
AP408   & 3 26 36.0   +49 36 58   & 21.392 & 2.719 & 0.043  0.258 &$>$16.1    --    &$>$15.6    --    &$>$15.3    --   &    --     --   &$>$18.02   --   & e       & N   \\  
AP409   & 3 20 17.7   +50 46 03   & 21.406 & 2.688 & 0.045  0.153 &$>$16.1    --    &$>$15.3    --    &$>$15.3    --   &   18.723 0.266 &   17.827 0.164 & a       & Y?  \\  
AP410   & 3 26 21.6   +48 44 50   & 21.565 & 2.766 & 0.060  0.519 &$>$16.1    --    &$>$15.6    --    &$>$15.3    --   &$>$18.37   --   &$>$17.5    --   & a       & Y?  \\  
AP411   & 3 26 10.1   +49 37 53   & 21.774 & 2.776 & 0.045  0.239 &$>$16.1    --    &$>$15.6    --    &$>$15.3    --   &    --     --   &   19.138 0.451 & e       & N   \\  
AP412   & 3 26 03.1   +49 30 29   & 22.057 & 2.952 & 0.071  0.342 &$>$16.1    --    &$>$15.6    --    &$>$15.3    --   &    --     --   &$>$18.60   --   & e       & N   \\  
\hline
\hline
\end{tabular}
}
$\,$\\  
a.- 1999 November. b 2000 February. c 2000 December. d 2001 November. e 2001 December \\
H data from CAHA.
\end{table*}


\begin{thebibliography}{}


\bibitem[2000]{ardila2000}
Ardila D.R., Mart\'{\i}n E.L., Basri G.,
2000, AJ 120, 479

\bibitem[1998]{baraffe98} 
Baraffe I., Chabrier G., Allard F., 
Hauschildt P. H.,
 1998, A\&A, 337, 403

\bibitem[1999]{ByN1999a}
Barrado y Navascu\'es D., Stauffer J.R., Patten, B.M., 
 1999, ApJ Letters 522, L56

\bibitem[2001]{barrado2001a} 
Barrado y Navascu\'es D., Stauffer J.R., Bouvier J., Mart\'{\i}n E.L., 
2001, ApJ 546, 1006 

\bibitem[2001]{barrado2001b} 
Barrado y Navascu\'es D., Stauffer. J.R., Brice\~no, C.,
 Patten B., Hambly N.C., Adams, J.D.,
2001, ApJ Suppl. 134, 103 

\bibitem[2002]{barrado2002} 
Barrado y Navascu\'es D., Stauffer. J.R., et al.
2002, in prep. 

\bibitem[1999]{basri1999} 
Basri G., Mart\'{\i}n E.L.,
1999, 510, 266

\bibitem[1996]{basri1996} 
Basri G., Marcy G.W., Graham J.R.,
1996, ApJ 458, 600

\bibitem[2000]{basri2000} 
Basri G., 
2000, Ann. Rev. A\&A 38, 485

\bibitem[2001]{bejar2001}
B\'ejar V., Mart\'{\i}n  E.L.,
 Zapatero Osorio M.R., Rebolo R.,
 Barrado y Navascu\'es D.,  Bailer-Jones C.A.L., Mundt R.,
 Baraffe I., Chabrier C., Allard F.,
2001, ApJ 556, 830

\bibitem[1998]{Bouvier1998}
Bouvier J., Stauffer J.R., Mart\'{\i}n, E.L., Barrado y Navascu\'es, D., 
Wallace B., B\'ejar, V., 
 1998, A\&A 336, 490

\bibitem[1998]{briceno1998}
Brice\~no C, Hartmann L., Stauffer J.R., Mart\'{\i}n E.L.,
1998, AJ 115, 2074

\bibitem[1997]{burrow1997} 
Burrow et al$.$ 1997, ApJ 491, 856

\bibitem[2000]{Chabrier2000}
Chabrier G.,  Baraffe I.,  Allard F., 
Hauschildt P., 
2000, ApJ,  542, L119.

\bibitem[1997]{DAntona1994}
D'Antona F., \& Mazzitelli I. 
1994 ApJ Suppl. 90, 467

\bibitem[1997]{DAntona1997}
D'Antona F., \& Mazzitelli I. 
1997, in ``Cool Stars in Clusters and 
Associations'', ed. R. Pallavicini \& G. Micela, Mem. Soc. Astron.
 Italiana, 68 (4), 807 

\bibitem[1997]{festin1997}
Festin L., 
1997, A\&A 322, 455

\bibitem[2000]{hodgkin2000}
Hodgkin S.T., Jameson R.F.,
2000, in ``Stellar Clusters and Associations:
Convection, Rotation, and Dynamos''.  
R. Pallavicini, G. Micela and S. Sciortino (eds.)
ASP Conf. Series 198, 245

\bibitem[1997]{jeffries1997} 
Jeffries R.D., Thurston M.R., Pye J.P.,
1997, MNRAS 287, 350

\bibitem[1998]{jeffries1998} 
Jeffries R.D., James D., J.,  Thurston M.R.,
1998, MNRAS 300, 550

\bibitem[2001]{jeffries2001} 
Jeffries R.D.,  Thurston M.R., Hambly N.C.,
2001, A\&A 375, 863

\bibitem[2002]{jameson2002} 
Jameson R.F.,
2002, in ``Brown Dwarfs'', IAU Symposium 211, eds. 
E.L. Mart\'{\i}n, in press

\bibitem[1996]{koester1996} 
Koester D., Reimers D.,
1996, A\&A 313, 810

\bibitem[1963]{kumar1963} 
Kumar S.S., 
1963, ApJ 137, 1121

\bibitem[1992]{Landolt1992}   
Landolt A., 1992, AJ 104, 340

\bibitem[1992]{Leggett1992}   
Leggett S., 1992 ApJ SS 82, 351

\bibitem[2000]{lucas2000}
Lucas P.W:, Roche P.F:, 
2000, MNRAS 314, 858

\bibitem[1999]{luhman1999}
Luhman K.L.,
1999, ApJ 525, 466 
 
\bibitem[2000]{luhman2000}
Luhman K.L., 
2000, ApJ 544, 1044

\bibitem[2000]{martin2000} 
Mart\'{\i}n E.L., Brandner W., Bouvier  J., Luhman K.L.,
 Stauffer, J., Basri G., Zapatero Osorio M.R.,  Barrado y Navascu\'es D.,
2000, ApJ 543, 299 

\bibitem[2001]{martin2001} 
Mart\'{\i}n E.L., Zapatero Osorio M.R.,  Barrado y Navascu\'es D.,
B\'ejar V.,  Rebolo R.,
2001, ApJ 558, 117

\bibitem[1993]{maynet1993} 
Meynet G., Mermilliod J.-C., Maeder A.,
1993, A\&A Series 98, 477

\bibitem[1992]{monet92} 
Monet D., et al.
 1992, AJ 103, 638.

\bibitem[2001]{moreaux2001}
Moraux E.,  Bouvier J, Stauffer J., 
2001, A\&A 367, 211

\bibitem[2000]{najita2000}
Najita J.R., Tiede G.P., Carr J.S.,
2000, ApJ 541, 977

\bibitem[1999]{neuhauser1999}
Neuh\"auser R., Comeron F., 
1999, A\&A 350, 612

\bibitem[2000]{pinfied2000} 
Pinfiled D.J., Hodgkin S.T., Jameson R.F., Cossburn M.R., Hambly N.C., Devereux N.,
2000, MNRAS 313, 347

\bibitem[1998]{pinsonneault1998} 
Pinsonneault M.H.,  Stauffer J., Soderblom D,R., King J.R., Hanson R.B.,
1998, ApJ 504, 170

\bibitem[1992]{prosser1992} 
Prosser C.F.,
1992, AJ 103, 488

\bibitem[1994]{prosser1994} 
Prosser C.F.,
1994, AJ 107, 1422

\bibitem[1996]{prosser1996}
Prosser  C.F., Randich  S.,
 Stauffer J.R., Schmitt J. H. M. M.,  Simon T.,
1996, AJ 112, 1570

\bibitem[1998]{prosser1998} 
Prosser C.F.,
1998, Astron. Nachr. 319, 215

\bibitem[1992]{rebolo1992} 
Rebolo R., Mart\'{\i}n E.L., Magazz\`u A.,
1992, ApJ Letters 389, 83

\bibitem[1996]{rebolo1996} 
Rebolo R., Mart\'{\i}n E.L., Basri G.,
 Marcy G.W., Zapatero-Osorio M.R.,
1996, ApJ 469, 706

\bibitem[1995]{rebolo1995} 
Rebolo R, Zapatero Osorio M.R., Mart\'{\i}n E.L., 
1995, Nature 377, 129

\bibitem[1985]{rieke1985}  
Rieke G.H.  \& Lebofsky M.J.,
1985, ApJ 288, 618

\bibitem[1999]{robichon1999} 
Robichon N., Arenou F., Mermilliod J.-C., Turon C.,
1999, A\&A 345, 471

\bibitem[1996]{santiago1996} 
Santiago  B.X., Gilmore G., Elson R.A.W.,
1996, MNRAS 281, 871


\bibitem[2000]{Siess2000}
Siess L., Dufour E., Forestini M., 2000, A\&A 358, 593

\bibitem[1997]{Skrutskie1997}  
Skrutskie  M. F., Schneider  S. E., Stiening  R.,
 Strom  S. E., Weinberg  M. D., Beichman  C.,
 Chester T., Cutri R., Lonsdale C., Elias J.,
 Elston  R., Capps R., Carpenter J., Huchra J.,
 Liebert J., Monet D., Price S., Seitzer P. 
 in ``The Impact of Large Scale Near-IR Sky Surveys'', 
 eds. F. Garzon et al., 
 p. 25. Dordrecht: Kluwer Academic Publishing Company, 1997

\bibitem[1985]{stauffer1985}
Stauffer  J. R., Hartmann  L. W.,
 Burnham  J. N., Jones  B. F.
1985  ApJ 289  247

\bibitem[1989]{stauffer1989}
Stauffer  John R.,
 Hartmann  Lee W., Jones  Burton F.
1989  ApJ 346  160

\bibitem[1998]{stauffer1998} 
Stauffer J.R.,  Schultz  G.,  Kirkpatrick J.D.,
1998, ApJ Letters 499, 199

\bibitem[1999]{stauffer1999} 
Stauffer J.R., Barrado y Navascu\'es D., Bouvier J., et al.
1999, ApJ 527, 219 

\bibitem[1986]{taylor1986}   
Taylor B.J.,
1986, ApJ Suppl.Series 60, 577

\bibitem[1998]{ventura1998} 
Ventura P., Zeppieri A., Mazzitelli I., D'Antona  F.,
 1998, A\&A 334, 953

\bibitem[1992]{wainscoat1992} 
Wainscoat R.J., Cohen M., Volk  K.,  Walker H.J., Schwartz D.E., 
1992, ApJ Suppl., 83, 111

\bibitem[1996]{zapatero1996} 
Zapatero Osorio M.R., Rebolo R.,   Mart\'{\i}n E.L., Garc\'{\i}a L\'opez R.,
1996, A\&A 305, 519

\bibitem[1997]{Zapatero1997}
Zapatero Osorio M.R., 
1997, PhD. Universidad de La Laguna, Spain

\bibitem[1999]{zapatero1999} 
Zapatero Osorio M.R., Rebolo R.,   Mart\'{\i}n E.L.,
 Hodgkin S.T.,  Cossburn M.R., Magazz\`u A., Stelle I.A., Jameson R.F.,
1999, A\&A Suppl. 134, 537 

\bibitem[2000]{zapatero2000} 
Zapatero Osorio M.R., B\'ejar V.J.S., Mart\'{\i}n E.L., Rebolo R.,   
Barrado y Navascu\'es D., Bailer-Jones C.A.L., Mundt, R., 
2000, Science 290, 103

\bibitem[2002]{Zapatero2002}
Zapatero Osorio  M.R., et al. 
2002, private communication.

%
   
\end{thebibliography}
\end{document}